# Study on flexible and organizable time-resolved measurement system and technology for multi-pulsed electron beam parameter[*]

Xiao-Guo Jiang(江孝国)1)　Yuan Wang(王远)　Guo-Jun Yang(杨国君)　Hong Li(李洪)　Zhuo Zhang(张卓)
Xing-Lin Yang(杨兴林)　Shu-Qing Liao(廖树清)　Tao Wei(魏涛)　Xiao-Ding Zhang(张小丁)　Yi-Ding Li(李一丁)
Institute of Fluid Physics，China Academy of Engineering Physics，Mianyang 621900，China

Abstract：The proof test and debugging of the multi-pulsed electron accelerator, Dragon-Ⅱ, requires a thorough comprehension of the quality of the beams. This puts forward a rigid requirement on the measurement system that it should have the ability that not only differentiates the three pulses on the whole but also tells the details of each pulse. In the measurements, beam energy is converted by a target to the Optical Transition Radiation (OTR) light, the information carried by which provides a good approach to measure beam profile and divergence simultaneously. Combining with this characteristic of OTR light, the concept of dual-imaging method is adopted in the design of optical imaging system. To avoid interference of the system optical parameters with one another, the original system is separated into two parts by functions, one for beam profile measurements and the other for divergence measurements. Correspondingly a splitter is interposed immediately after the OTR target which splits the light into two parts in perpendicular directions, one part forming a beam spot, and the remaining becoming polarized through a polarizer and imaged by an OTR lens and forming a beam divergence pattern. Again the use of splitters helps the light beam enter different framing camera for different use. With a time resolution of 2ns, the system can provide 8 frames of beam details within any one of the three pulses, at the same time as acquiring the quality of the three pulses on the whole.

Keywords：multi-pulsed electron beam, beam emittance, time-resolved measurements, optical transition radiation, dual-imaging method

PACS：41.85.Ja；41.85.Ew；29.27.Eg

## 1 Introduction

The development of multi-pulsed electron beam linear induction accelerators(LIA) which helps to provide data with higher veracity and efficiency by obtaining multi-axis and multi-frame images in one explosion has become more and more compulsory for flash x-rayradiography in hydrodynamic test. For this purpose, high performance LIADARHT-I[1] and DARHT-II[2,3] have been developed in LANL. DARHT-I produces one pulse and DARHT -II produces up to four pulses in a variable pulse format at one time. These two LIA are at right angles to each other. We have also designed and constructed two LIA, Dragon-I[4] and Dragon-II[5], which are arranged perpendicularly to each other and work together. Dragon-I produces one pulse and Dragon-II produces up to three pulses at one time. Succeeded to set up of Dragon-I in 2003, Dragon-II has been put into application in 2015. It can produce three successive electron beam pulses in one shot within about 1μs and the duration of each pulse is about 70ns with intervals of about 500 ns.

The measurement work of multi-pulsed electron beam parameters is very complex and very difficult. It may be break into two parts of between pulses and inner pulse in time-resolved character. The former means that the parameters should be distinguished from each other pulses and this means that it is ok if integral parameters in whole pulse duration or instantaneous parameters at any time of one pulse are obtained. The latter means that the parameters should be distinguished temporally even in the same pulse and this means that the parameters is time dependent in one pulse. In order to meet these measurement demands, many studies have been carried out on new testing facilities development such as super high speed three-frame framing camera[6], eight-frame framing camera[7] and a variety of beam parameter measurement systems[8,9,10] have also been developed in recent years. For Dragon-II, some new beam parameters measurement should be developed to meet the complex time-resolved demands and to improve debugging efficiency. In this paper, we designed a new electron beam parameters measurement system which can meet many kinds of time-resolved demands for the multi-pulsed beam

Referenced on 1 May 2016, Revised on 1 May 2016
*Supported by the National Natural Science Foundation of China (10675104,11375162)
1) Jiang Xiao Guo(1968-), Master of engineering, works on the technique of time-resolved measurements of high-current electron beam parameters.
E-mail:j_xg_caep@sina.com

measurements. This system is flexible and organizable. It can be assembled easily according to the measurement aim at work field.It helps to acquire athorough understanding of the multi-pulsed electron beams property and has expedited electron beams debugging work.

## 2　Time sequences designed for multiple time-resolved measurements of Dragon-II

For the 3-pulse electron beams generated in Dragon-II, time-resolved measurements requires the measuring system to resolve the three successive pulses from each other and to detect instantaneous status of the beam within each pulse at any time.The resolution between pulses can be further divided into two aspects, integral detections for any one pulse but absolutely distinguished from each other, see Fig.1(a),and instantaneous detections for any one pulse but at the same corresponding time in its duration,see Fig.1(b).The inner-pulse measurement can provide 8 instantaneous detections for any one of the three pulses at one time, see Fig.1(c).$\tau_0$ shown in Fig.1 is the delay time before the shutter opens after the camera is triggered, $\tau$ is the camera's shutter time, and $\tau_i$ is the interval time between two exposures. In high-speed mode, $\tau_0$ and $\tau_i$ can be set in step of smaller than 1ns respectively, and $\tau$ can be set to 2ns to the extreme. These performance helps to improve the synchronicity of the measurement and enhance the relevance[11]. The data measured or calculated to some extent reveals the beam's actual state at that time.The system can fulfill a series of complicate tests and measurements for Dragon-II because of its flexibility and combinations.

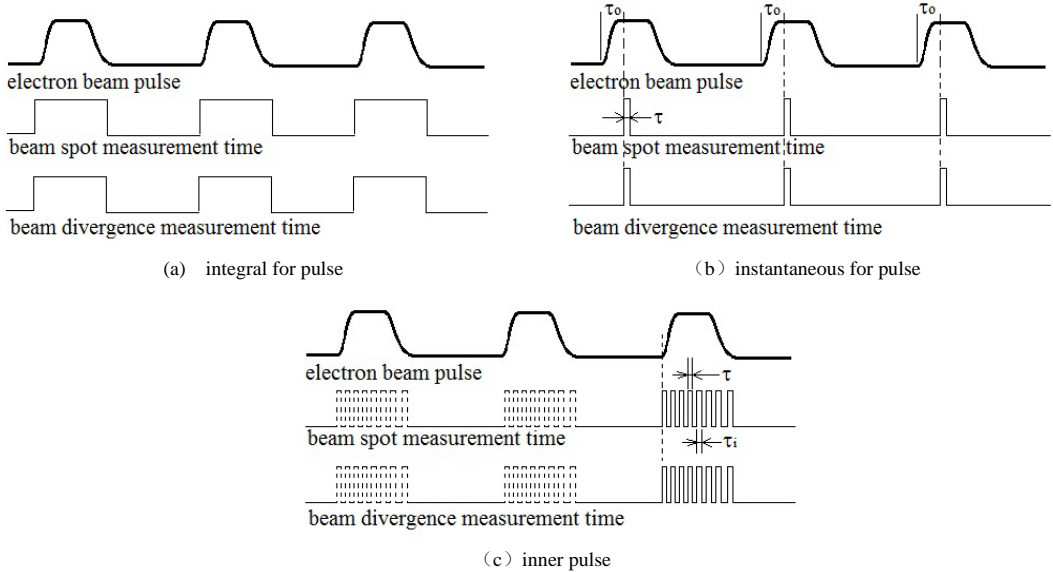

Fig.1. Time sequences for the measurements of the Dragon-Ⅱ electron beam characteristics

## 3　Optical principle for the measuring system

In the measurements, Optical Transition Radiation (OTR) light is employed to measure beam profile and divergence.OTR light is optical light given off by a charged particle when it gets across the interface of two mediums. It has characteristic pattern of angular distribution, with its radiated energy centering about the direction of $1/\gamma$. Here $\gamma$ is the relativistic energy factor of the particle.The measuring system is assembled on the basis of dual-imaging method[12], see Fig.2. It helps to classify the lights from the object space according to its position on the image plane, and to its direction on the focal plane at the same time. Thus images on image plane and focal plane respectively carry out the information about beam profile and divergence.

Successfully applied to the development of the instantaneous emittance measuring system, the dual-imaging principle combined with OTR method is not ready for immediate utilization in experiments. There are still two difficulties to overcome. First, despite concision and directness, the dual imaging function of the lensin this principle layout means restriction from freely adjustment for the two factors we concerned.And also the lack of space is a main constraint of multi spot measurements. Fortunately, we have an alternate design which is functionally separated and thus more practical to meet the requirement of multi-spot independent measurements of beam, see Fig.3.

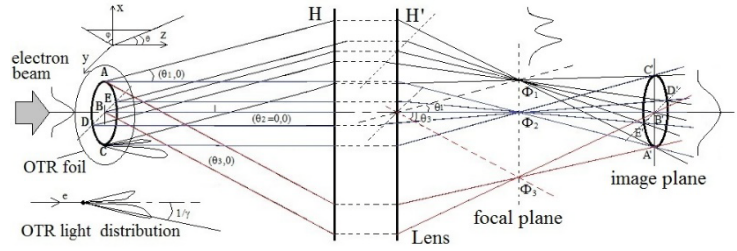

Fig.2. OTR method and dual-imaging principle

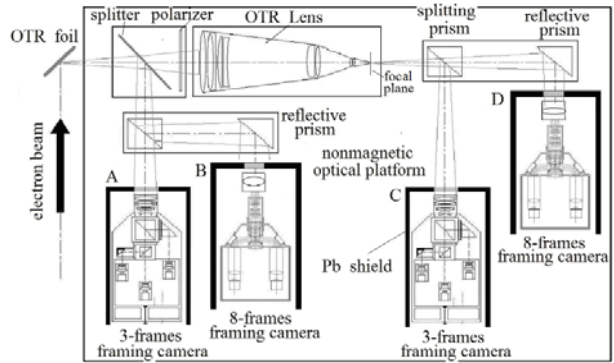

Fig.3. Layout of the separate function system

When the light from OTR foil hits the splitter, it is split into two parts which enter different light paths, one transmitted ahead for divergence measurement use and the other reflected in the perpendicular direction for measurement of profile. Then they both are split again and accepted by a corresponding framing camera. We can see from Fig.3 that camera A and B record the image on OTR foil, and camera C and D record that on the focal plane of OTR lens, with imaging parameters of the optic systems in camera A, B, C and D being independent from each other. Profit from the separation of divergence and profile and the independence of recording system, this system can fulfill a variety of measuring demands through different combination of subassemblies.

For the 18MeV electron beam in Dragon-II, the focal length of the OTR lens is 450mm, the aperture is 280mm, and the distance form the OTR lens's forehead to the OTR foil should be within 1.2m.

## 4    Experimental research of the time-resolved beam parameters measuring system

The whole time-resolved measurement system for 3 pulsed electron beams by Dragon-II is shown in Fig.4.

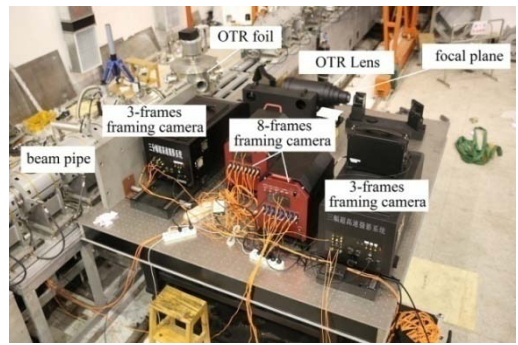

Fig.4. Fieldwork layout of the measurement system

Fig.5 shows the typical images for beam spot and divergence under distinguishing mode between pulses. Fig.5(a),(b) and (c) are spot image. Fig.5(d),(e) and (f) are divergence image. The camera's shutter time of each exposure is 150ns, with a time lag of 500ns between two successive exposures.

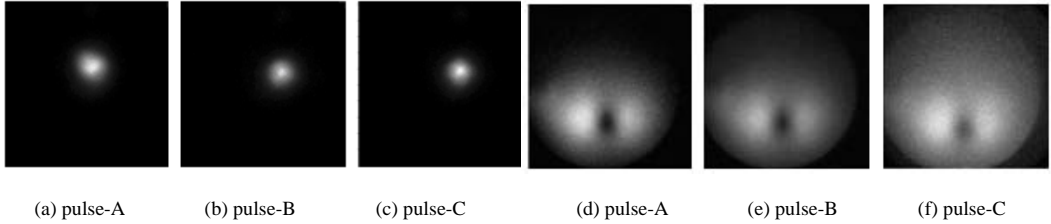

(a) pulse-A     (b) pulse-B     (c) pulse-C     (d) pulse-A     (e) pulse-B     (f) pulse-C

Fig.5. Electron beam images for each pulse under distinguishing mode between pulses

The successive images captured in one pulse are shown in Fig.6. The camera's shutter time is 10ns and the interval time is 10ns. The fourth frame image is left blank in spot image sequence because the corresponding camera channel is destroyed.

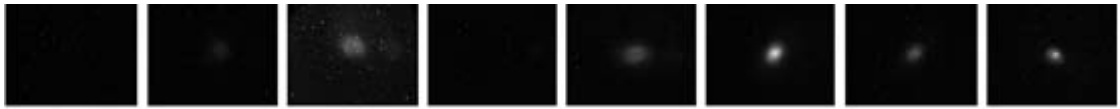

(a) Images of beam spots

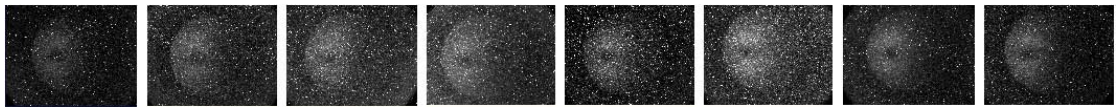

(b) Images of beam divergence

Fig.6. Time-resolved beam spots and divergence images within one pulse

## 5 Results of the typical beam parameters measurements

Beam parameters are calculated from the measured images. Fig.7 shows the data read from beam spot image of pulse C in Fig.5(c) and its Gaussian fitting curve. Read out the image gray degree along a centroidal axis of the spot and calculate the FWHM value along this direction. Repeat the above procedure in a series of directions that cover the whole imaging plane and take the average as the final FWHM value of beam profile.

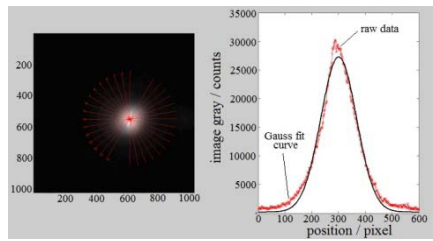

Fig.7 Data process for beam size

For the divergence image, draw a line through the two centers where the peak energy located, and read out the image gray degree along this line. Beam energy measured and optic parameters of the OTR lens known, we can calculate a set of standard curves for a given series of divergence distribution. Compare these curves to the gray degree curve, and we can say that the divergence distribution corresponding to the curve the closest to the experimental data to some extent represents that of the beam, as shown in Fig.8.

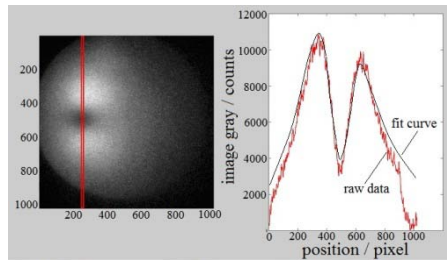

Fig.8. Data process for beam divergence

Table 1 gives the results of differentiationof pulse from each other. Table 2 is the time-resolved measurements within one pulse.

Table 1 Integral parameters of each pulse

| Pulse NO | Beam Energy /MeV | Spot Radii /mm | Beam divergence /mrad |
|---|---|---|---|
| A | 18.85 | 13.3 | 4.0 |
| B | 18.60 | 11.0 | 4.1 |
| C | 18.85 | 10.5 | 4.5 |

Table 2 Time-resolved parameters of one pulse

| Serial NO | Beam Energy /MeV | Spot Radii /mm | Beam divergence /mrad |
|---|---|---|---|
| 1 | 18.60 | 12.3 | 4.5 |
| 2 | 18.60 | 14.7 | 4.0 |
| 3 | 18.60 | 16.0 | 3.7 |
| 4 | 18.60 | / | 3.8 |
| 5 | 18.60 | 15.9 | 3.7 |
| 6 | 18.60 | 8.9 | 5.9 |
| 7 | 18.60 | 8.1 | 6.3 |
| 8 | 18.60 | 7.5 | 6.6 |

## 6  Discussion

As a method helping to recognize the state of the multi-pulse electron beam on the whole, the beam parameter measuring system in Fig.4 is very useful in beam diagnose and debugging for its comprehensive adaptability to meet the various demands of the measuring tasks of Dragon-II. Change the combination of the components, and you can get a derivative system for a new use.  Fig.9 shows a system especially designed for beam spot measurements, simple and effective. Fig.10 shows a combined system which is used in the proof-test of Dragon-II injector. In this system, complementary to the framing camera, a streak camera is adopted for beam envelope evolution measurement. This combination helps to make a thorough reorganization of the transport quality of the injector segment[13].

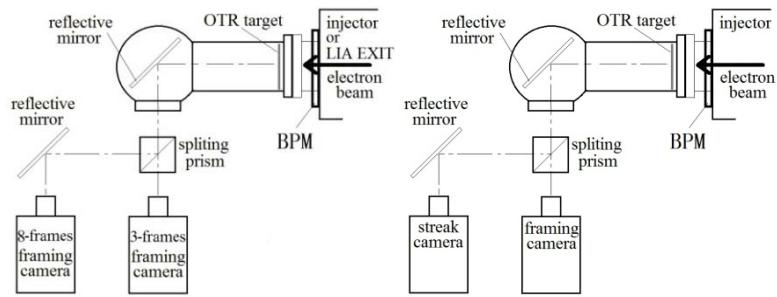

Fig.9. Layout of the time-resolved spot measurementsystem Fig.10Layout of the measurement system of beam spot and envelope

## 7  Conclusion

Ameasurement system is designed for the time-resolved detection of beam parameters of the multi-pulse electron beam in Dragon-II LIA. With this system, beam profile and divergence can be recorded simultaneously. Flexibility in combination of components makes it a system adaptable to various measuring demands of electron beam in Dragon-II.As a surveillance equipment of beam, it greatly improvesthe efficiency of the facility's proof-test process.